\def\lsim{\lower -0.3ex \hbox{$<$} \kern -0.75em \lower 0.7ex \hbox{$\sim$}}
\def\gsim {\raise.35ex\hbox{$>$}\kern-0.75em\lower.5ex\hbox{$\sim$}}
\def\beq{\begin{equation}}
\def\eeq{\end{equation}}
\def\beqa{\begin{eqnarray}}
\def\eeqa{\end{eqnarray}}

\def\ie{${\rm i.e.}$}
\def\etal{${\it et\: al.}$}
\def\subbeqa{\begin{subeqnarray}}
\def\subeeqa{\end{subeqnarray}}

\def\Aro{A_{\rho}}
\def\Cro{C_{\rho}}

\newcommand{\dk} {{\rm d}k}

\newcommand{\dq} {{\rm d}q}
\newcommand{\dt} {{\rm d}t}
\newcommand{\dx}{{\rm d}x}
\newcommand{\dy}{{\rm d}y}

\documentstyle[preprint,eclepsf]{jpsj}
\def\runauthor{Michiyasu {\sc Mori}, Masao {\sc Ogata} and Hidetoshi {\sc Fukuyama}}

\title
{
Quantized Conductance of One-Dimensional Doped Mott Insulator
}

\author
{ 
Michiyasu {\sc Mori}$^{1,2}$, Masao {\sc Ogata}$^3$\\
and \\
Hidetoshi {\sc Fukuyama}$^2$
}

\inst
{
$^1$ Institute for Molecular Science, Okazaki 444, Japan\\ 
$^2$ Department of Physics, the University of Tokyo, 7-3-1 Hongo, 
Bunkyo-ku, Tokyo 113\\
$^3$ Department of Basic Science, Graduate School of Arts 
and Sciences,\\
the University of Tokyo, Komaba, Meguro-ku, \\
Tokyo 153, Japan}

\recdate
{
\today
}

\abst{The possible modification of quantized conductance of one-dimensional doped Mott insulator, where the Umklapp scattering plays an important role, is studied based on the method by Maslov-Stone and Ponomarenko. At $T=0$ and away from half-filling, the conductance is quantized as $g=2e^2/h$ and there is no renormalization by Umklapp scattering process. At finite temperatures, however, the quantization is affected depending on the gate voltage and temperature.}

\kword
{
conductance, quantization, two terminal measurement, reserviors, Mott transition, Umklapp scattering process, bosonization, quantum sine-Gordon model, massive Thirring model
}

\begin{document}
\sloppy
\maketitle

With recent progress in the micro-fabrication techniques, it has become possible to design not only point contacts but also quantum wires.\cite{garcia,tarucha,pfeiffer} For transport properties of  quantum wires, we expect interesting phenomena due to finite size of the system, impurity scattering and mutual interaction. Far away from half-filling, one-dimensional interacting electron system behaves as Tomonaga-Luttinger liquid which has different properties from the conventional Fermi liquid. As for the conductance of Tomonaga-Luttinger liquid, the experiment by Tarucha \etal\cite{tarucha} negated the existence of the possible modification of the quantized value as, $g=(2e^2/h)K_{\rho}$,\cite{apel,ogatafukuyama} where $K_{\rho}$ is a correlation exponent determined by interaction and electron density.\cite{kawakami,korepin,schulz} Actually more recent studies have disclosed that the quantization of conductance is not changed by interaction, if only the effect of leads\cite{maslov,ponomarenko,safi,shimizu,finkel} 
or the renormalization of the electric field  by the interaction\cite{kawabata} are taken into account. In addition to that, it has been clarified that the quantized conductance is affected  by disorder.\cite{ogatafukuyama,maslov2}

In the Tomonaga-Luttinger regime, the interaction is restricted to processes with small transfer, \ie, forward scatterings, leading to no excitation gap. Then a question arises: What happens if the large momentum transfer process, \ie, backward or Umklapp scattering process, becomes important. Since the backward scattering does not essentially change the elementary excitation of charge degrees of freedom at least in long systems, the effect of Umklapp scattering, which becomes important near the half-filling, is of great interest. Actual realization of such situation seems to be possible in a quantum wire modulated with periodic potential  which is originally proposed by Ogata and Fukuyama\cite{ogatafukuyama2} and is fabricated very recently by Tarucha\cite{tarucha2} in split gate structure grown on GaAs with extra periodic potential along the wire. 

\begin{figure}[h]
\begin{center}
\epsfile{file=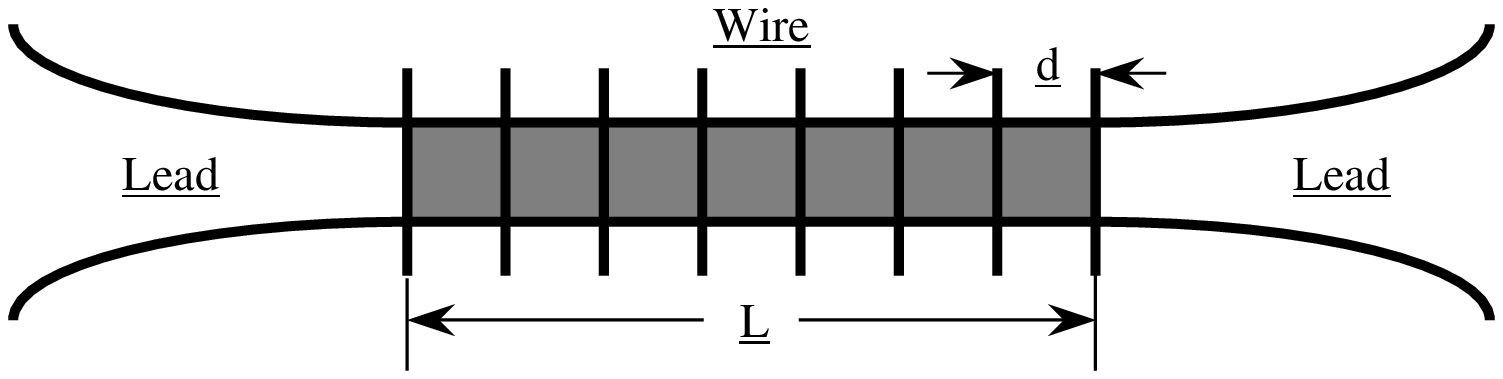,height=4cm,width=12cm}
\end{center}
\caption{The quantum wire modulated with periodic potential.}
\label{modqwire}
\end{figure}
\begin{figure}[b]
\begin{center}
\epsfile{file=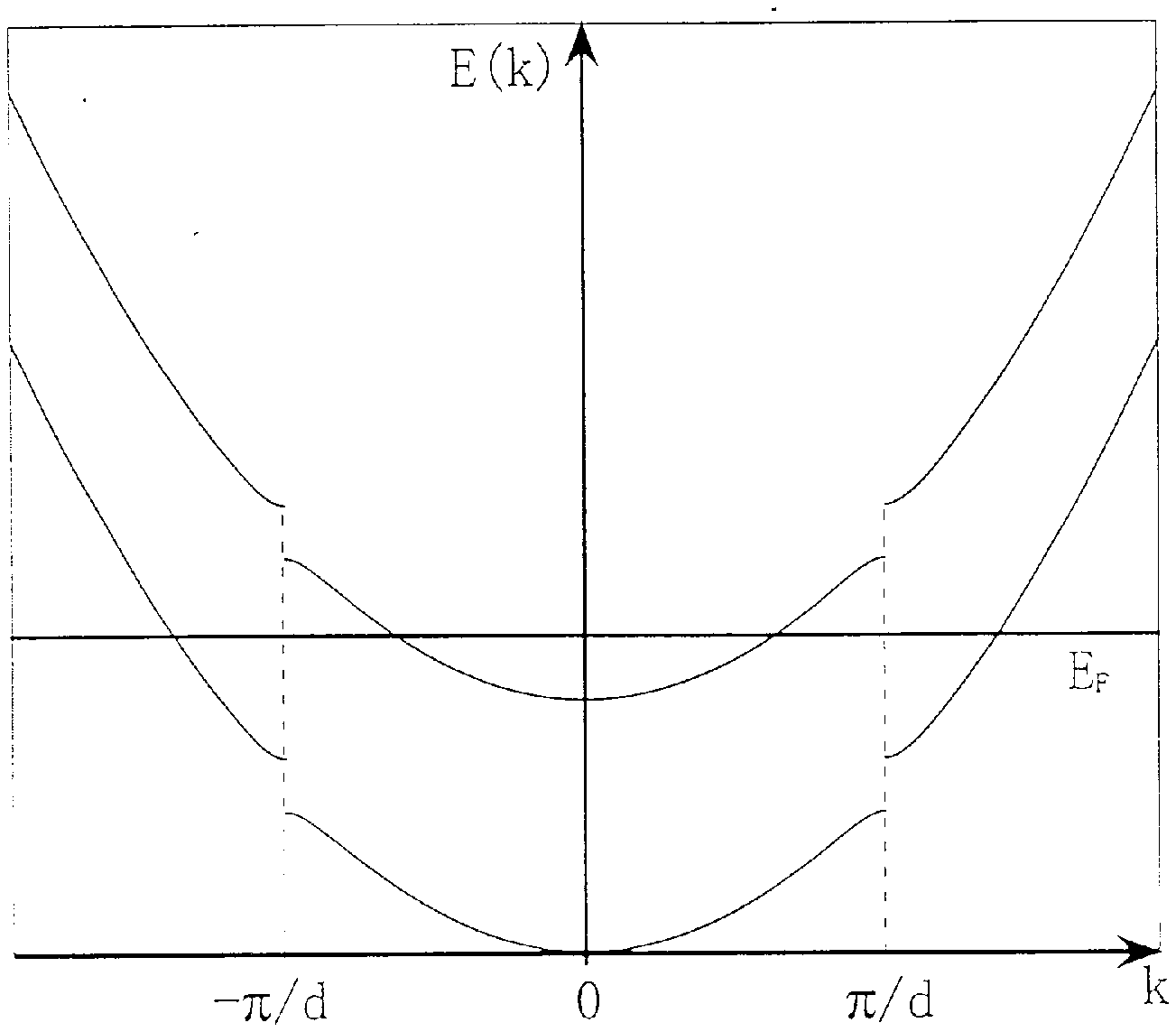,height=6cm}
\end{center}
\caption{The sub-bands accompanied with band gap.}
\label{subband}
\end{figure}
The introduction of extra periodic potential is schematically shown in Fig.\ref{modqwire}. This will make some channels (the second channel in Fig.\ref{subband}) half-filled bands, if the gate voltage controlling the Fermi level is properly chosen. 
In the presence of Coulomb interaction, the half-filled channel will be a Mott insulator because of the Umklapp scattering process. It is to be noted that the Mott-Hubbard gap at half-filling is always present in one-dimensional systems,\cite{liebwu} once the interaction is finite. Hence the way how the conductance of such a channel varies as a function of the gate voltage, \ie, in the region of the doped Mott insulator, is of particular interest. In this paper, we will make it clear whether the mutual interaction including Umklapp scattering  process renormalizes the quantized value of conductance or not. Although some authors\cite{fujimoto,ponomarenko2} have treated the Umklapp process perturbatively, we develop an alternative method to study the conductance which enables us to take account of the non-perturbative effect of Umklapp process. At zero temperature, we find that even the Umklapp scattering as far as away from half-filling does not renormalize the conductance,
while at $T \neq 0$ the quantized value is affected as a function of 
temperature and gate voltage.   

To calculate the dc conductance of two-terminal measurement, we take into account  leads which are attached to both ends of quantum wire.\cite{shimizuueda} We suppose that leads are 'continuously' connected to the quantum wire. Formally, the word 'continuous' means that the Green function and its derivative for the long wavelength fluctuation of charge density are continuous at the interfaces between the 'wire' and 'leads'. Physically it means that there is no charge accumulation at the interfaces and then the current is conserved. The Coulomb interaction in the leads can be neglected, since the width of the leads, $W_{lead}$, is sufficiently large compared to that of wire, $W_{wire}$, \ie, $W_{lead}\gg W_{wire}$. Thus we now treat the whole system as an inhomogeneous one-dimensional system,\cite{maslov,ponomarenko,safi} which is composed of the interacting quantum wire ($0\leq x\leq L$) and the non-interacting leads ($x < 0, L < x$) continuously connected to the wire. In addition to the above condition, the inter-band scattering and the reflection at the interfaces between the wire and leads are neglected.     

\begin{figure}[t]
\begin{center}
\epsfile{file=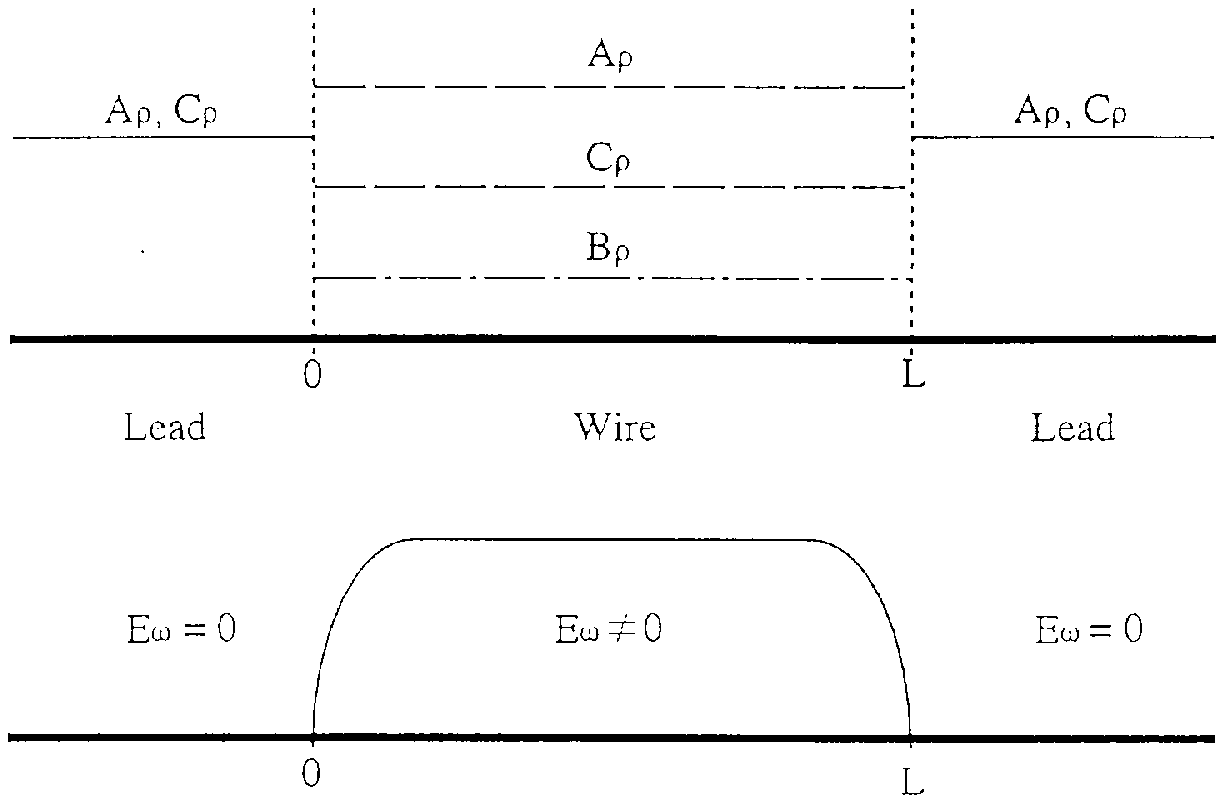,height=7cm}
\end{center}
\caption{Spatial variation of $\Aro(x)$, $B_{\rho}(x)$, $\Cro(x)$ and 
${\rm E}_{\omega}(x)$ in the two-terminal measurement. $B_{\rho}(x)$ is the 
coupling constant of Umklapp scattering.  $L$ is the system size.}
\label{parameter}
\end{figure}

By taking into account such an inhomogeneity of the whole system, we adopt the following effective Hamiltonian for the low energy excitations of charge degrees of freedom, which is derived in bosonization,\cite{suzumura,solyom,emery,fukuyama}
\beq
{\cal H}_B = \int\dx \left\{\Aro(x)(\nabla \theta(x))^2 
+ B_{\rho}(x)\cos(2\theta(x) - q_0 x) + \Cro(x)P(x)^2 \right\},
\label{phase}
\eeq
where 
\beq
[\theta(x),P(y)]={\rm i}\delta(x-y),\nonumber
\label{commu}
\eeq
 and $\theta(x)$ describes the fluctuation of charge degrees of freedom. Here, $B_{\rho}(x)$-term originates from the Umklapp scattering\cite{emery2,emery3,giamarchi,giamarchi2,giamarchi3,morifukuima,mori} which is  finite inside the 'wire' and  disappears in the 'leads' as shown in Fig.\ref{parameter}.
In eq.(\ref{phase}), $q_0=G-4k_0$ is the misfit parameter where $G$ is the reciprocal lattice vector, $G=2\pi/d$, with the period, $d$,   
and $k_0=\mu/v_{F0}$ is the Fermi momentum with $\mu$ and $v_{F0}$ 
being the chemical potential and the Fermi velocity of the non-interacting 
electrons, respectively. 
Other kinds of interactions are renormalized into the parameters $\Aro(x)$ and $\Cro(x)$. 

We calculate current response linear to the electric field which is 
finite in the 'wire' and zero in the 'leads', 
as shown in Fig.\ref{parameter}.  
Thus the following relations is obtained: 
\beqa
j(x,\omega)
&=&- \int_{0}^{L}\dy {\rm E}_{\omega}(y) \int_{-\infty}^{\infty}\dt 
\frac{e^{i\omega t} -1}{i\omega} K^R(x,y;t),\nonumber\\
&=& \int^{L}_{0}\dy {\rm \sigma}(x,y ; \omega) {\rm E}_{\omega}(y),
\label{kuboformula}
\eeqa
where  $K^R(x,y;t)$ is the retarded current-current correlation function. 

Since the current operator, $j(x)$, in the bosonization scheme is given by,
\beqa
j(x)
&=&-\frac{e}{\pi}\dot{\theta}(x),\\
&=&-e\frac{2C_{\rho}}{\pi}P(x),
\eeqa
and $\theta(x)$ couples to the external electric field, ${\rm E}_{\omega}(x)$, as, 
\beq
H_{ext}=\frac{e}{\pi}\int_{0}^{L}\dy \theta(y){\rm E}_{\omega}(y)e^{i\omega t},
\label{extfield}
\eeq
we have following relations, 
\beqa
K^R(x,y;t)
&=& -(\frac{e}{\pi})^2 \frac{i}{\hbar}\Theta(t)\langle[\dot{\theta}(x,t),
\dot{\theta}(y,0)]\rangle,\label{acourse}\\
&=&-(C_{\rho}(x))(C_{\rho}(y))(\frac{2e}{\pi})^2
\frac{i}{\hbar}\Theta(t)\langle[P(x,t),P(y,0)]\rangle,\nonumber\\
\label{bcourse}
\eeqa
where $\Theta(t)$ is the step function. 
Using this current-current correlation function,  
$\sigma(x,y; \omega)$ is rewritten as,
\beqa
\sigma(x,y;\omega)
&=& i\omega(\frac{e}{\pi})^2 G^R(x,y;\omega), \label{asigma}\\
&=&\frac{-1}{i\omega}(\frac{2e}{\pi})^2 (C_{\rho}(x))(C_{\rho}(y))
\left\{ Q^R(x,y;\omega) - Q^R(x,y;0) \right\},\nonumber\\
\label{bsigma}
\eeqa
in terms of a retarded Green function for $\theta$, 
\beq
G^R(x,y;\omega)
\equiv \int_{-\infty}^{\infty}\dt e^{{\rm i}\omega t}
\frac{-i}{\hbar}\Theta(t)\langle[\theta(x,t),\theta(y,0)]\rangle\\
\label{greenfn}
\eeq
 and  a correlation function, 
\beq
Q^R(x,y;\omega)
\equiv  \int_{-\infty}^{\infty}\dt e^{{\rm i}\omega t}
\frac{-i}{\hbar}\Theta(t)\langle[P(x,t),P(y,0)]\rangle.\\
\label{ppcorre}
\eeq

In the Tomonaga-Luttinger regime where $B_{\rho}(x)=0$, Maslov-Stone\cite{maslov} and Ponomarenko\cite{ponomarenko} calculated the conductance by solving the equation of motion for $G^R(x,y;\omega)$ in the inhomogeneous system and showed that the conductance is not renormalized. 
Near half-filling, however, the $B_{\rho}$-term plays an important role and the equation of motion for $G^R(x,y;\omega)$ can not be solved in a similar way because the Umklapp scattering process gives the following non-linear term,  
\beq
2B_{\rho}\left(\frac{{\rm -i}}{\hbar}\right)\Theta(t)
\langle \left[\sin(2\theta(x,t) - q_0 x),\theta(y,0)\right]\rangle.
\eeq

In order to resolve this difficulty, we adopt the Luther-Emery's method\cite{luther} to the charge degrees of freedom\cite{emery3,giamarchi,morifukuima,mori} and map the bosonized Hamiltonian, eq.(\ref{phase}), onto the following Hamiltonian of spinless Fermion,
\begin{eqnarray}
H_{F}  
 & = & \int {\rm d}x v_c(x)(\Psi^{\dag}(x) (- {\rm i}\partial\tau_{3})\Psi(x)) 
+ \frac{v_c(x) q_0}{2} \Psi^{\dag}(x)\Psi(x)\nonumber\\
 & + &\int {\rm d}x  V(x)\Psi^{\dag}(x) \tau_{1} \Psi(x)\nonumber\\
 & + & \int {\rm d}x \frac{W(x)}{2/\pi}[(\Psi^{\dag}(x)  \Psi(x))^2
-(\Psi^{\dag}(x) \tau_{1} \Psi(x))^2], 
\label{thirring}
\end{eqnarray} 
where 
\beqa
\Psi(x)
&=&
\left(\begin{array}{c}
                \psi_{1}(x) \\ \psi_{2}(x)
                \end{array}\right), \nonumber\\
v_c(x)\nonumber
& = &\pi A_{\rho}(x)+C_{\rho}(x)\frac{1}{\pi}
=v_{\rho}(x)\left(\frac{1}{4 K_{\rho}(x)}+ K_{\rho}(x)\right),\\
V(x)\nonumber
& = & B_{\rho}(x)(\pi\alpha),\\
W(x)\nonumber
& = &\pi A_{\rho}(x)-C_{\rho}(x)\frac{1}{\pi}
=v_{\rho}(x)\left(\frac{1}{4 K_{\rho}(x)}- K_{\rho}(x)\right),\\
\label{mapping2}
\eeqa
$v_{\rho}(x)=2\sqrt{A_\rho(x) C_\rho(x)}$ is the velocity of charge excitation and $\tau_j$ (${\it j}$=0,1,2,3) are Pauli matrices. 
$V(x)$ is identified with the Umklapp scattering which is finite in the 'wire' and zero in the 'leads' as shown in Fig.\ref{parameter}.
Since $K_{\rho}$ approaches 1/2 near the half-filling, \cite{kawakami,korepin,schulz} the assumption that $W(x)=0$ in the 'wire' but $W(x)\neq0$ in the 'leads' is a good approximation. 
In terms of spinless Fermion, the conjugate field of the phase variable, $\theta(x)$, 
is expressed as, 
\beq
P(x)=2\Cro(x)(\Psi^{\dag}(x)\tau_3 \Psi(x)).
\label{identity}
\eeq
Here, we derive $G^R(x,y;\omega)$ of the doped Mott insulator by calculating  $Q^R(x,y;\omega)$ and then using identities eq.(\ref{asigma}) and (\ref{bsigma}). 
After obtaining the solutions for $G^R(x,y;\omega)$ inside the wire and the leads, independently, $G^R(x,y;\omega)$ of the whole system is determined by use of boundary conditions at interfaces between wire and leads in the spirit of Maslov-Stone\cite{maslov} and Ponomarenko\cite{ponomarenko}. 

\begin{figure}[h]
\begin{center}
\epsfile{file=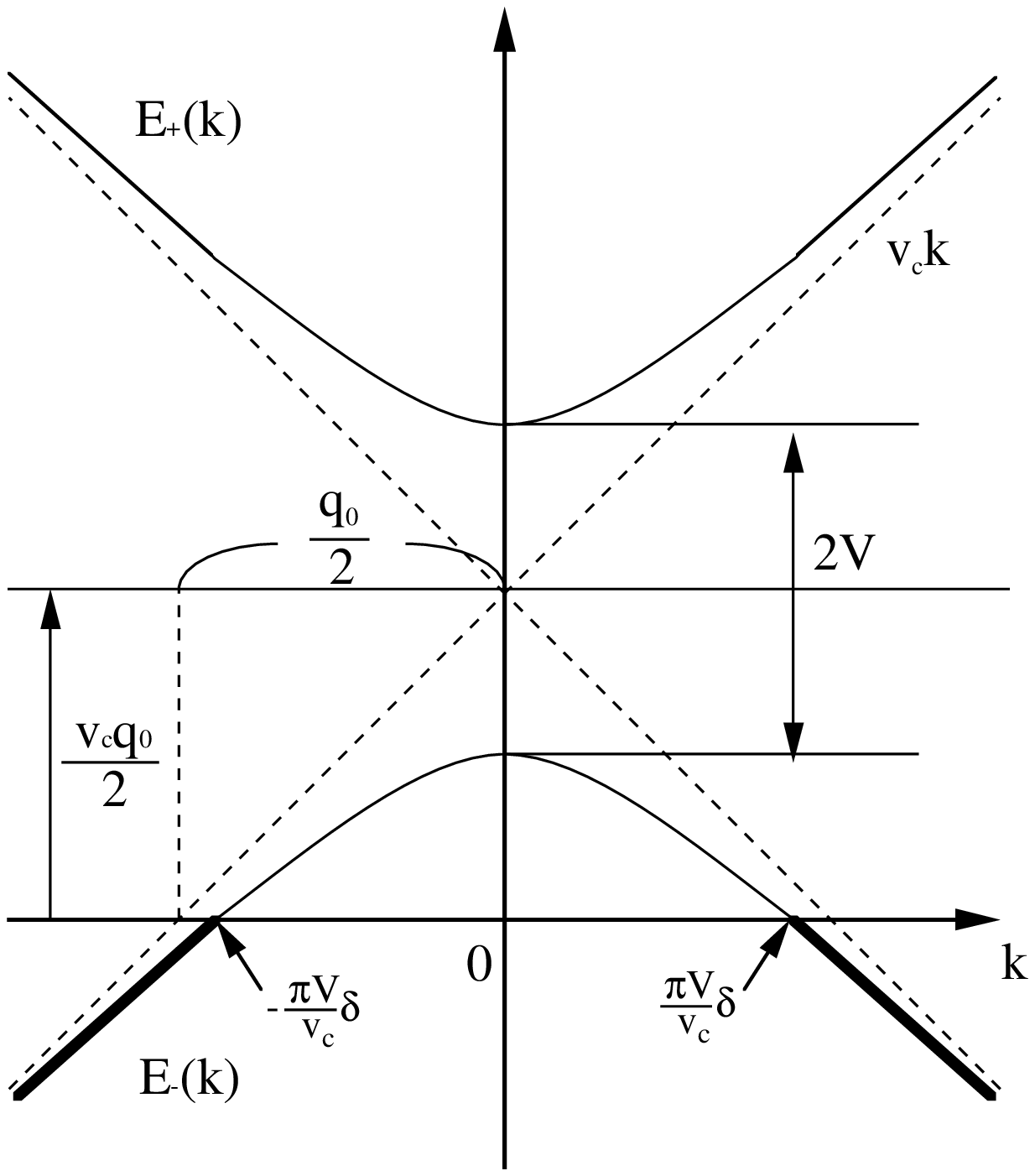,height=8cm,width=10cm}
\end{center}
\caption{The elementary excitation for the charge degree of freedom.}
\label{dispersion}
\end{figure}

First, we calculate the correlation function, $Q^R(x' ,y ; \omega)$, inside the wire  for the following region of $\omega$ and $T$, 
\beq
0\leq\omega, T \ll \frac{v_c q_0}{2}-V,
\label{lowt}
\eeq
where $v_c q_0 /2 -V$ is the energy difference between the Fermi energy 
and the top of lower Hubbard band as shown in Fig.\ref{dispersion}. Here, $x'$ and $y$ are restricted into the 'wire'.
By taking into the above conditions, $Q^R(x' ,y;\omega)$ is calculated as follows, 
\beqa
&&Q^R(x',y; \omega)\nonumber\\
&=& \int\frac{\dq}{2\pi}e^{{\rm i}q(x'-y)} \Big[ 
\nonumber\\
&&\left. \int\frac{\dk}{2\pi} M_{k,q} 
\left\{\frac{f(v_c q_0 /2-E_k) - f(v_c q_0 /2-E_{k+q})}{\omega - E_k + E_{k+q} + {\rm i}\eta} 
+ \frac{f(v_c q_0 /2+E_k) - f(v_c q_0 /2+E_{k+q})}{\omega - E_{k+q} + E_k + {\rm i}\eta } \right\}\right.\nonumber\\
&+& \int\frac{\dk}{2\pi} N_{k,q} 
\left.\left\{ \frac{f(v_c q_0 /2-E_k) - f(v_c q_0 /2+E_{k+q})}{\omega - E_k - E_{k+q}+ {\rm i}\eta } 
+ \frac{f(v_c q_0 /2+E_k) - f(v_c q_0 /2-E_{k+q})}{\omega + E_k + E_{k+q}+ {\rm i}\eta } \right\}\right] \nonumber\\
&&\label{qfn1}
\eeqa
where $f(\epsilon)$ is the Fermi distribution function and 
\beqa
E_k &=& \sqrt{(v_c k)^2 + V^2},\nonumber\\
\left.
\begin{array}{rr}
u_k\\
v_k
\end{array}
\right\}
&=&
\frac{1}{2}(1 \pm \frac{v_c k}{E_k}),\nonumber\\
M_{k,q}
&=& 
u_{k+q}u_k + v_{k+q}v_k - \frac{V^2}{2E_{k}E_{k+q}},\nonumber\\
N_{k,q}
&=& 
u_{k+q}v_k + v_{k+q}u_k + \frac{V^2}{2E_{k}E_{k+q}}.
\eeqa

Since we are interested in the limit of $\omega\rightarrow0$, the third and fourth term in eq.(\ref{qfn1}) can be neglected within the first order of $\omega$. If the temperature is lower than the Mott-Hubbard gap, moreover, the second term is of relative magnitude $\exp \{- \beta(v_c q_0 / 2 + V)\}$ to the first term and then can be ignored. Thus $Q^R(x',y; \omega)$, eq.(\ref{qfn1}), is approximated as,
\beq
Q^R(x',y;\omega) \sim \int\frac{\dq}{2\pi}e^{{\rm i}q(x'-y)}\int\frac{\dk}{2\pi} M_{k,q} 
\frac{f(v_c q_0 /2-E_k) - f(v_c q_0 /2-E_{k+q})}{\omega - E_k + E_{k+q}+ {\rm i}\eta }.
\label{qfn2}
\eeq
It is seen that $q\sim0$ is dominant in the limit of $\omega \rightarrow 0$.  
Therefore, by expanding eq.(\ref{qfn2}) as for $q$ to the first order, 
we obtain $G^R(x',y;\omega)$ in the wire based on $Q^R(x',y;\omega)$ as follows, 
\beqa
G^R(x',y ;\omega)
&\sim& \frac{(2C_\rho)^2}{2 \pi {\rm i}\omega}\left[
\left\{1-f(\frac{v_c q_0}{2} - V)\right\}\right.\nonumber\\
&&\left.\left\{\theta(x'-y)\exp\left({\rm i}\frac{\omega}{v_g}(x'-y)\right) + 
\theta(y-x')\exp\left({\rm -i}\frac{\omega}{v_g}(x'-y)\right)\right\} + {\cal O}(\omega)\right],\nonumber\\
\label{gfnmott}
\eeqa
where 
\beq
v_g \equiv v_c \sqrt{1-(2V/v_c q_0)^2}\sim \pi v_c \delta, \nonumber
\eeq
with $\delta$ being given by,
\beq
\delta =\frac{1}{\pi V}\sqrt{(\frac{ v_c q_0}{2})^2-V^2}.
\eeq
Here, $v_g$ and $\delta $ are the group velocity at the Fermi energy in the lower Hubbard band and the doping rate, respectively. 
The expansion as regards $q$ in eq.(\ref{qfn2}) to arrive at eq.(\ref{gfnmott}) is valid in the following region, 
\beq
\frac{v_c}{V}\frac{1}{\delta} \ll |x'-y| < L,
\label{cond1}
\eeq
where the first inequality is obtained from the condition that the first order in the expansion of $E_{k+q}$ is large enough in comparison with the second order.

Next, $G^R(x'',y;\omega)$ in the 'leads' are obtained as follows by considering that the 'leads' are the non-interacting systems,
\beqa
G^R(x'',y;\omega)&=&\left\{ 
\begin{array}{ll}
\exp(-i\frac{\omega}{v_F}(x''-y)), \hspace{1cm}(x''< 0), \\
\exp(+i\frac{\omega}{v_F}(x''-y)), \hspace{1cm}( L<x''). \nonumber\\
\end{array}
\right.
\eeqa

Finally, by taking into account the above results, the solution of the each region in the limit of 
$\omega\rightarrow0$ is given by
\beqa
G^R(x,y;\omega)&=&\left\{ 
\begin{array}{ll}
A \exp(-i\frac{\omega}{v_F}x), \hspace{1cm}(x< 0), \\
B \exp(-i\frac{\omega}{v_g}x)+C \exp(+i\frac{\omega}{v_g}x), 
\hspace{1cm}(0\leq x < y\leq L), \\
D \exp(-i\frac{\omega}{v_g}x)+E \exp(+i\frac{\omega}{v_g}x),  
\hspace{1cm}(0\leq y < x\leq L), \\
F \exp(+i\frac{\omega}{v_F}x), \hspace{1cm}( L<x), \nonumber\\
\end{array}
\right.\\
\eeqa
where $v_F$ is the Fermi velocity of non-interacting system. $A\sim F$ are functions  of $y, \omega$ and $T$ determined by the boundary conditions. 

At $T=0$, by following the same procedure used by Maslov-Stone\cite{maslov} and Ponomarenko,\cite{ponomarenko} we can conclude that the conductance of doped Mott insulator is $g = 2e^2/h$ if the mean distance between holes, $1/\delta$, is small enough compared to the size of system, $L$, \ie, eq.(\ref{cond1}).
(At half-filling, \ie, Mott insulator, however, since the current-current correlation function is always zero in the 
limit of $\omega\rightarrow0$, 
the conductance should be zero.) 
For $T\neq0$, on the other hand, the quantized conductance  deviates from the universal value as , 
\beq
\left\{1-f(\frac{v_c q_0}{2}-V)\right\}\frac{2e^2}{h}.
\label{tempdep}
\eeq
The above results will be valid for, 
\beq
{\rm Max}\left\{\frac{\hbar v_c}{k_B T}, \frac{v_c}{V}\frac{1}{\delta}\right\}
\ll
L
\ll
\frac{v_g}{\omega}\sim (\pi v_c)\frac{\delta}{\omega},
\eeq
where the discreteness of energy level caused by the finite size of system is ignored by the first inequality. The second inequality always holds when we consider the dc limit. 
Here, it is noted that, as $1/\delta$ becomes comparable to $L$, our treatment seems  to break down and may crossover to a different picture.

Although, at $T=0$, there is no difference between the Tomonaga-Luttinger liquid and doped Mott insulator, the Umklapp process becomes relevant for $T \neq 0$ and the conductance decreases by increasing the temperature or approaching the Mott insulator. The temperature dependence in eq.(\ref{tempdep}) is valid only for the low temperature region, \ie, eq.(\ref{lowt}).
For higher temperature as, $T \gsim v_c q_0 /2 -V$, it is hard to calculate $ G^R(x',y;\omega)$ since the expansion of $E_{k+q}$ with respect to $q$ is not a  good approximation. 
If we extrapolate this result to such higher temperature region, the conductance would be reduced to a value $g=e^2/h$ since $f(0)=1/2$.  Although this extrapolation is not valid, the result seems to make sense, \ie, the charge degrees of freedom in the lower 
Hubbard band is described by spinless Fermions and the spin summation in 
the Landauer formula drops out, which indicates the spin-charge separation 
in the vicinity of Mott transition.  
When temperature is higher than Mott-Hubbard gap, we expect that the upper Hubbard band begins to contribute to the transport and the universal value of conductance ($g=2e^2/h$) will be recovered. 

Recently, a few studies have been carried out on the effect of Umklapp scattering on the 
quantized conductance.\cite{fujimoto,ponomarenko2} 
Their studies, however, are based on the perturbative calculation with respect to 
the strength of the Umklapp term. 
Since the renormalization group study shows that the Umklapp scattering renormalizes
to the strong coupling near the half-filling, the perturbative calculation is not valid. 
On the contrary, our method using the mapping to the spinless Fermion model takes 
account of the non-perturbative effect of the Umklapp scattering. 
Thus, we can discuss the effect of Mott-Hubbard gap explicitly in our formulation and conclude that the quantization does not collapse near the Mott transition at 
zero temperature, in contrast to Fujimoto and Kawakami.\cite{fujimoto}

In summary, 
we have studied the quantized conductance of doped Mott insulator by the two terminal measurement. 
Such situation can be realized in the quantum wire modulated with 
periodic potential where 
some particular channels approach the half-filling by varying the gate voltage and the Umklapp scattering process plays an important role. 
The conductance of such a case is calculated based on the method by Maslov-Stone and Ponomarenko. 
At $T=0$ and away from half-filling, the conductance is always $2e^2/h$ as far as  the mean distance of holes is smaller than the size of system. However for $T\neq0$, the quantized conductance of doped Mott insulator deviates from the universal value as, $(1-f(v_c q_0 / 2-V))(2e^2 / h)$ at low temperatures, $T \ll v_c q_0 / 2 -V.$

We would like to thank Professor A. Kawabata, Professor N. Nagaosa and Professor A. Shimizu for useful discussions. This work was financially supported by a Grant-in-Aid for Scientific 
Research on Priority Area "Anomalous Metallic State near the Mott 
Transition" (07237102), and ``Novel Electronic States in Molecular Conductors'' (06243103) from the Ministry of Education, Science, Sports and Culture.


\end{document}